\documentclass[aps,showpacs,superscriptadress,preprint]{revtex4}
\usepackage{graphicx}

\begin{document}
\title{Liquid-gas Phase Transition in Strange Hadronic Matter}

\author{L. Yang$^1$, W.L. Qian$^1$\footnote{wlqian@fudan.edu.cn},
R.K. Su$^{2,1,4}$\footnote{rksu@fudan.ac.cn} and H.Q.
Song$^{2,3}$\footnote{songhq@sinr.ac.cn;Ref:24-430}} \affiliation{
\small 1. Department of Physics, Fudan University,Shanghai 200433, China\\
\small 2. CCAST(World Laboratory), P.O.Box 8730, Beijing 100080, China\\
\small 3. Shanghai Institute of Applied Physics, Chinese Academy
of Sciences, P.O.Box 800204, Shanghai 201800, China \\
\small 4. Center of Theoretical Nuclear Physics, National
Laboratory of Heavy Ion Collisions, Lanzhou 730000, China}

\begin{abstract}
The liquid-gas phase transition in strange hadronic matter is
studied utilizing an extended Furnstahl-Serot-Tang model with
nucleons and hyperons.  The system is treated as of two
components.  The phase transition is analyzed by investigating the
stability of the system and Gibbs conditions for phase
equilibrium.  A two-dimensional binodal surface resulted from the
two-phase equilibrium, namely the phase boundary, is obtained. For
each temperature ranging from $T = 8 MeV$ to $T = 12 MeV$, a limit
pressure on the binodal surface section is found, while a critical
point is spotted  for the temperature around $T = 13 MeV$.  The
Maxwell constructions are also illustrated to give a vivid
description of the course of the phase transition. Moreover, the
entropy per baryon and heat capacity per baryon as functions of
temperature are examined.  The entropy is continuous during the
phase transition but the heat capacity is discontinuous,
indicating that the phase transition is of second order.  By these
efforts, the L-G phase transition can be concluded to exist in the
strange hadronic matter.
\end{abstract}

\maketitle

\section{Introduction}

Since the nuclear interaction is very similar to the Van der Waals
potential between molecules, the possible existence of
liquid-gas(L-G) phase transition in nuclear matter was predicated
by theoretical studies\cite{Kupper,Lamb,Bertsch}.  This
predication was supported later by intermediate-energy heavy-ion
collisions\cite{Curtin} and high-energy proton-induced
reactions\cite{Finn}.
 The idea of an L-G phase transition in nuclear matter emerged and attracted much interest\cite{Jaqaman,Su,Satpathy,Kapusta,Goodman,Boal,Siemens}.
 This interest has increased in the last decade, with the attempts by
the EOS Collaboration to extract exponents of 1 GeV/nucleon Au
nuclei with a carbon target\cite{EOS} and with the extraction by
the ALADIN/LAND Collaboration of a caloric curve resulting from
the fragmentation of the quasiprojectile formed in the collision
Au+Au at 600 MeV/nucleon exhibiting a behavior expected for a
first-order L-G phase transition\cite{ALADIN}. Very rich
information has already been extracted from exclusive experimental
data during the intermediate energy heavy ion collisions, as the
new generation $4\pi$ detectors have been developed and are now
operating at different accelerator facilities, such as Dubna,
GANIL, GSI, LNL, LNS, MSU and Texas A-M\cite{Agostino}.

This L-G phase transition in nuclear matter occurs at lower
density than the normal nuclear density $\rho_{0}$
\cite{nucl-th/0305050}.  The two phases are of cold nuclear Fermi
liquid on the one hand and a nuclear gas consisting of free
nucleons on the other hand.  For symmetric nuclear matter, having
equal number of protons and neutrons, this L-G phase transition is
known of first order.  However, M\"{u}ller and Serot have
indicated that the phase transition in asymmetric nuclear matter
is of second order by Ehrenfest's definition\cite{Muller}.  The
entropy and the volume as the first order derivatives of the
chemical potentials are continuous and the heat capacity as the
second order derivative is discontinuous at the point of phase
transition.  The phase coexistence surface of this binary system
of two phases is two dimensional\cite{Muller, Qian1, Qian2, Qian3,
Qian4, Kolomietz, Karnaukhov, Pawlowski}.

Since the first hypernucleus was detected in emulsion by Danysz
and Pniewski\cite{Danysz} in 1953, strangeness carried by s-quark
has opened a new dimension for the studies in nuclear physics.
With the hyperons included in, nuclear matter is extended to
strange hadronic matter(SHM).  Employing different models, the
equation of state and the stability of the SHM have been
investigated\cite{Ikeda, Schaffner1, Schaffner2, Schulze, Lee,
Zhang1}.  However, besides the deconfinement phase transition to
the quark gluon plasma(QGP), whether there exists an L-G phase
transition in the SHM is still an open question in theoretical
physics. As the interactions between hyperon and nucleon or
between hyperons are also of the behavior of the Van der Waals
potential, it is reasonable to guess that there will also be L-G
phase transition in the SHM.  It is the main purpose to study this
problem.

It is believed that the descriptions of nuclear matter and finite
nuclei are ultimately governed by the physics of low-energy
quantum chromodynamics(QCD).  Because of the nonperturbative
properties of QCD, it is very difficult to establish a fundamental
theory of nuclear system from QCD.  One usually adopts various
phenomenological models based either on hadron degree of freedom
or quark degree of freedom.  In our previous papers\cite{Zhang1,
Qian5}, we suggested an effective model, constructed by
introducing hyperons in the Furnstahl-Serot-Tang(FST)
model\cite{Furnstahl1997, Furnstahl1995}, to study the saturation
properties and stabilities of the SHM.  The details of this model
can be found in ref.\cite{Zhang1}.  Here we only give a short
description.

In refs.\cite{Zhang1, Qian5}, we extended the original FST model
to include not only nucleons and $\sigma$, $\omega$ mesons, but
also $\Lambda$, $\Xi$ hyperons.  A new hyperon-hyperon interaction
mediated by two additional strange mesons $\sigma^*$ and $\phi$
,which couple only to hyperons, is introduced. The reactions
$\Lambda + \Lambda \rightarrow \Xi^{-} + p$, $\Lambda + \Lambda
\rightarrow \Xi^{0} + n$ and their reverses are considered.  In
this model, we did not consider the mixture of the ${\Sigma}$
hyperons. The reason, as was explained in ref.\cite{Zhang1}, is
twofold. First, the ${\Sigma}$ potential in nuclear matter at the
saturation density is rather uncertainly predicted, ranging from
completely unbound\cite{Mares} to $U_{\Sigma} = -25\pm 5
MeV$\cite{Dover}. As pointed out by Balberg et al\cite{Balberg},
systems involving ${\Sigma}$'s together with nucleons or
$\Lambda$'s generally will be unstable with respect to the strong
decays $\Sigma + N \rightarrow \Lambda + N$ or $\Sigma + \Lambda
\rightarrow \Xi + N$.  Secondly, the Q values for the strong
transitions $\Sigma + N \rightarrow \Lambda + N$, $\Sigma + \Sigma
\rightarrow \Lambda + \Lambda$, $\Sigma + \Lambda \rightarrow \Xi
+ N$ and $\Sigma + \Xi \rightarrow \Lambda + \Xi$
 are about $78, 156, 50$ and
$80 MeV$, respectively\cite{Stoks}.  To Pauli-block these
processes, we need a rather high density of $\Lambda$.  On the
other hand, the Q value of $\Xi + N \rightarrow \Lambda + \Lambda$
is only about $28 MeV$.  This process can be Pauli-blocked by a
relatively low $\Lambda$ density.  Based on this model, the
stability, the density, temperature and strangeness fraction
dependence of the effective masses of nucleons and baryons, the
pressure, the free energy and the equation of state are studied,
and the results are reasonable.

The paper evolves from an attempt to study the L-G phase
transition of SHM by employing the extended FST model.  In Sec. II
the extended FST model is laid out.  The SHM is modelled into a
two-component system ready for the discussion of phase transition.
The thermodynamic quantities for this model are briefly derived
out. By analyzing the equilibrium conditions the phase sections
are addressed in Sec. III.  The thermodynamical properties and the
phase transition are discussed and the order of the phase
transition is also examined.  Sec. IV finally summarizes the main
results about the existence of the L-G phase transition in the
SHM.

\section{The extended FST model}

The original FST model is extended by including $\Lambda$ and
$\Xi$ hyperons and two additional strange mesons $\sigma^{\ast}$
and $\phi$ to describe the hyperons and the interactions between
them as suggested in ref.\cite{Zhang1}.  Since we are studying the
unpolarized system, the $\pi$ meson has no influence on the system
and is omitted.  To reduce the degrees of freedom, we need some
assumptions.  First, we assume $\Xi ^{-}$ and $\Xi ^{0}$ appear in
equal amount.  Also we have protons and neutrons in equal amount,
which is similar to protons and neutrons in symmetric nuclear
matter.  These assumptions mean that we are looking at matter with
zero isospin.  But the remarkable new degree of freedom,
strangeness is included.  The lagrangian of the extended FST model
is presented as follows:
\begin{eqnarray}
{\cal L}(x) &=&\bar{\psi}_{N}(i\gamma ^{\mu }\partial _{\mu
}-g_{\omega N}\gamma ^{\mu}V_{\mu}-M_{N}+g_{sN}\sigma )\psi _{N}
\nonumber \\ &&+\bar{\psi}_{\Lambda }(i\gamma ^{\mu }{\partial
}_{\mu }-g_{\omega \Lambda }\gamma ^{\mu}V_{\mu}-g_{\phi \Lambda
}\gamma ^{\mu}\phi _{\mu}-M_{\Lambda }+g_{s\Lambda }\sigma
+g_{\sigma ^{\ast }\Lambda }\sigma^{\ast })\psi _{\Lambda }
\nonumber \\ &&+\bar{\psi}_{\Xi }(i\gamma ^{\mu }{\partial }_{\mu
}-g_{\omega \Xi }\gamma ^{\mu}V_{\mu}-g_{\phi \Xi }\gamma
^{\mu}\phi _{\mu}-M_{\Xi }+g_{s\Xi }\sigma+g_{\sigma ^{\ast }\Xi
}\sigma^{\ast })\psi _{\Xi }  \nonumber \\
&&-\frac{1}{4} G_{\mu\nu}G^{\mu\nu} +\frac{1}{2}\left( 1+\eta
\frac{\sigma }{S_{0}}\right) m_{\omega
}^{2}V_{\mu}V^{\mu}+\frac{1}{4!}\zeta \left( g_{\omega
N}^{2}V_{\mu}V^{\mu}\right)^{2}\nonumber \\
&&+\frac{1}{2}\partial_{\mu}\sigma\partial^{\mu}\sigma
-H_{q}\left( \frac{S ^{2}}{S_{0}^{2}}\right) ^{2/d}\left[
\frac{1}{2d} ln\left( \frac{S ^{2}}{S_{0}^{2}}\right)
-\frac{1}{4}\right]\nonumber \\
&&-\frac{1}{4}S_{\mu\nu}S^{\mu\nu}+\frac{1}{2}m_{\phi }^{2}\phi
_{\mu}\phi^{\mu}+\frac{1}{2}\left(\partial_{\nu}\sigma^{\ast
}\partial^{\nu}\sigma^{\ast }-m_{\sigma ^{\ast }}^{2}\sigma ^{\ast
^{2}}\right)
\end{eqnarray}
In  mean-field approximation, the lagrangian can be written as
\begin{eqnarray}
{\cal L}_{MFT} &=&\bar{\psi}_{N}(i\gamma ^{\mu }\partial _{\mu }
+g_{\omega N}\gamma ^{0}V_{0}-M_{N}+g_{sN}\sigma _{0})\psi _{N}
\nonumber \\ &&+\bar{\psi}_{\Lambda }(i\gamma ^{\mu }{\partial }
_{\mu }-g_{\omega \Lambda }\gamma ^{0}V_{0}-g_{\phi \Lambda }
\gamma ^{0}\phi _{0}-M_{\Lambda }+g_{s\Lambda }\sigma _{0}
+g_{\sigma ^{\ast }\Lambda }\sigma _{0}^{\ast
})\psi _{\Lambda }  \nonumber \\ &&+\bar{\psi}_{\Xi }(i\gamma ^{\mu }
{\partial }_{\mu }-g_{\omega \Xi }\gamma ^{0}V_{0}-g_{\phi \Xi }
\gamma ^{0}\phi _{0}-M_{\Xi }+g_{s\Xi }\sigma _{0}+g_{\sigma ^{\ast }
\Xi }\sigma _{0}^{\ast })\psi _{\Xi }  \nonumber \\ &&+\frac{1}{2}
\left( 1+\eta \frac{\sigma _{0}}{S_{0}}\right) m_{\omega }^{2}V_{0}^{2}
+\frac{1}{4!}\zeta \left( g_{\omega N}V_{0}\right) ^{4}+\frac{ 1 }{2}
m_{\phi }^{2}\phi _{0}^{2}-\frac{1}{2}m_{\sigma ^{\ast }}^{2}\sigma ^{\ast ^{2}}
  \nonumber \\ &&-H_{q}\left( 1-\frac{\sigma _{0}}{S_{0}}\right) ^{4/d}
  \left[ \frac{1}{d} ln\left( 1-\frac{\sigma _{0}}{S_{0}}\right) -\frac{1}{4}\right]
\end{eqnarray}
where the meson field operators have been replaced by their mean
field values: $\phi _{0}$, $V_{0}$ , $\sigma _{0}$ and $\sigma
_{0}^{\ast }$.  $ g_{ij}$ are the coupling constants of baryon $j$
to meson $i$ field.  By using the standard technique of
statistical mechanics, the thermodynamic potential $\Omega $ is
obtained
\begin{eqnarray}
\Omega &=&V\{H_{g}[(1-\frac{\sigma
_{0}}{S_{0}})^{\frac{4}{d}}(\frac{1}{d} \ln (1-\frac{\sigma
_{0}}{S_{0}})-\frac{1}{4})+\frac{1}{4}]  \nonumber \\
&&-\frac{1}{2}(1+\eta \frac{\sigma _{0}}{S_{0}})m_{\omega
}^{2}V_{0}^{2}-  \frac{1}{4!}\zeta (g_{\omega
N}V_{0})^{4}-\frac{1}{2}m_{\phi }^{2}\phi
_{0}^{2}+\frac{1}{2}m_{\sigma ^{\ast }}^{2}\sigma _{0}^{\ast
^{2}}\}  \nonumber \\ &&-2k_{B}T\{\sum_{i,{\bf k}}\ln
{[1+e^{-\beta (E_{i}^{\ast }(k)-\nu _{i}}] } +\sum_{{i,{\bf
k}}}\ln {[1+e^{-\beta (E_{i}^{\ast }(k)+\nu _{i})}]}\}
\end{eqnarray}
where $\beta \ $is the inverse temperature and $V$ is the volume
of the system.
\begin{equation}
E_{i}^{\ast }(k)=\sqrt{M_{i}^{\ast 2}+k^{2}}
\end{equation}
with the effective masses of the hyperons and nucleons
\begin{equation}
M_{i}^{\ast }=M_{i}-g_{si}\sigma _{0}-g_{\sigma ^{\ast }i}\sigma
_{0}^{\ast }\qquad(i=\Lambda ,\Xi ),
\end{equation}
\begin{equation}
M_{i}^{\ast }=M_{i}-g_{si}\sigma _{0}\qquad(i=N)
\end{equation}
The mean-field values $\phi _{0}$, $V_{0}$ , $\sigma _{0}$ and
$\sigma _{0}^{\ast }$ are determined by the corresponding extreme
conditions of the thermodynamic potential.  For instance,the
equation for $\phi $ meson is determined by
\begin{equation}
\frac{\partial \Omega }{\partial \phi _{0}}|_{\sigma _{0},V_{0},\sigma
_{0}^{\ast },\mu _{i}}=0
\end{equation}
in explicit form
\begin{equation}
m_{\phi }^{2}\phi _{0} - g_{\phi \Lambda }\rho _{B\Lambda
}-g_{\phi \Xi }\rho _{B\Xi }=0.
\end{equation}
The  baryon densities $\rho _{Bi}$ is given by
\begin{equation}
\rho _{Bi}=\left\langle {\psi}^{+}_{i}\psi _{i}\right\rangle
=\frac{g_{i}}{ \pi ^{2}}\int dkk^{2}\left[ n_{i}\left( k\right)
-\bar{n}_{i}\left( k\right) \right]
\end{equation}
where $g_{i}=4$ for $i=N$ or $\Xi $, $g_{i}=2$ for $i=\Lambda $.
The baryon and anti-baryon distributions are, respectively,
expressed as
\begin{equation}
n_{i}(k)=\{exp[\beta (E_{i}^{\ast }(k)-\nu _{i})]+1\}^{-1}
\end{equation}
and
\begin{equation}
\overline{n}_{i}(k)=\{exp[\beta (E_{i}^{\ast }(k)+\nu _{i})]+1\}^{-1}
\end{equation}
in which $\nu _{i}$ are related to chemical potential $\mu _{i}$ by
\begin{eqnarray}
\mu _{N} &=&\nu _{N}+g_{\omega N}V_{0},  \nonumber \\
\mu _{\Lambda } &=&\nu _{\Lambda }+g_{\omega \Lambda }V_{0}+g_{\phi \Lambda
}\phi _{0},  \nonumber \\
\mu _{\Xi } &=&\nu _{\Xi }+g_{\omega \Xi }V_{0}+g_{\phi \Xi }\phi
_{0}.
\end{eqnarray}
In the system with equal number of protons and neutrons as well as
equal number of $\Xi^0$ and $\Xi^-$, the chemical equilibrium
condition for the reactions $\Lambda +\Lambda \rightleftharpoons n
+\Xi^0$ and $\Lambda +\Lambda \rightleftharpoons p +\Xi^-$ reads
\begin{equation}
2 \mu_\Lambda=\mu_N +\mu_\Xi.
\end{equation}
Eq.(13) implies that only two components are independent among
$N$, $ \Lambda $ and $ \Xi $.  The system of SHM described by the
extended FST model is a two-component system.  The nucleons ($p$
and $n$) and the hyperons ($ \Lambda $, $ \Xi^{-} $ and $ \Xi^{0}
$) play the role of different components, respectively. The
strangeness fraction is introduced as
\begin{equation}
 f_{S}\equiv
\frac{\rho _{B\Lambda }+2\rho _{B\Xi }}{\rho _{B}}
\end{equation}
which plays the similar role as that of the asymmetric parameter $
\alpha = ( \rho_{n} - \rho_{p}) / (\rho_{n} + \rho_{p})$ in the
asymmetric nuclear matter.  We can use the same method as that in
ref.\cite{Muller,Qian1,Qian2,Qian3,Qian4} to address the L-G phase
transition.

Following the usual procedure of statistical physics, we can
easily calculate the other thermodynamic quantities from
thermodynamic potential $\Omega $.  For example, the pressure and
entropy density are calculated by formulas $p=-\Omega /V$ and $
S/V=-(\partial \Omega /\partial (1/\beta ))_{V,\mu
_{i}}/V=(\partial p/\partial (1/\beta ))_{V,\mu _{i},\sigma
_{0},V_{0},\phi _{0},\sigma _{0}^{\ast }}$.  The results are
expressed as follows
\begin{eqnarray}
p &=&\sum_{i}\frac{g_{i}}{6\pi ^{2}}\int dk\frac{k^{4}}{E_{i}^{\ast }(k)}
[n_{i}(k)+\overline{n}_{i}(k)]-H_{q}\left\{ \left( 1-\frac{\sigma _{0}}{S_{0}
}\right) ^{\frac{4}{d}}\left[ \frac{1}{d}ln\left( 1-\frac{\sigma _{0}}{S_{0}}
\right) -\frac{1}{4}\right] +\frac{1}{4}\right\}   \nonumber \\
&&+\frac{1}{2}\left( 1+\eta \frac{\sigma _{0}}{S_{0}}\right)
m_{\omega }^{2}V_{0}^{2}+\frac{1}{4!}\zeta g_{\omega
N}^{4}V_{0}^{4}+\frac{1}{2} m_{\phi }^{2}\phi
_{0}^{2}-\frac{1}{2}m_{\sigma ^{\ast }}^{2}\sigma _{0}^{\ast
^{2}},
\end{eqnarray}
\begin{eqnarray}
s &\equiv &S/V=\sum_{i}\frac{g_{i}}{6\pi ^{2}}\int dk\frac{k^{4}}{
E_{i}^{\ast }(k)}\left\{ \frac{\beta ^{2}(E_{i}^{\ast }\left( k\right) -\nu
_{i})\exp \left[ \beta \left( (E_{i}^{\ast }(k)-\nu _{i})\right) \right] }{
\left[ \exp \left[ \beta \left( (E_{i}^{\ast }(k)-\nu _{i})\right) \right] +1
\right] ^{2}}\right.   \nonumber \\
&&\left. +\frac{\beta ^{2}(E_{i}^{\ast }\left( k\right) +\nu
_{i})\exp \left[ \beta \left( (E_{i}^{\ast }(k)+\nu _{i})\right)
\right] }{\left[ \exp \left[ \beta \left( (E_{i}^{\ast }(k)+\nu
_{i})\right) \right] +1\right] ^{2}}. \right\}
\end{eqnarray}

\section{L-G phase transition}
In this section, we employ the extended FST model to investigate
the L-G phase transition in the SHM at different strangeness
fractions.  In the calculation, the parameter set T1 given in
ref.\cite{Zhang1} is used.

As the hot and dilute SHM is only obtained on earth by the
relativistic heavy ion collision, the investigations of the
thermodynamic properties of the SHM are based on two basic
assumptions.  One is that we can apply equilibrium thermodynamics
for such a small system of only few hundred constituents at the
most.  The other is that a thermalized uniform system is formed in
heavy ion collision before the multi-fragmentation takes
place\cite{Lee2}.  Although the equilibrium analysis
oversimplifies the study of the SHM, we still follow the
thermodynamic approach in the reason that this can give some
concrete descriptions of the phase structure of the SHM and
characterize certain aspects of the evolution.

We would like to discuss the stability of SHM first.  To address
the stability, we consider the Helmholtz free energy density
${\cal F}(T,\rho_i)$ defined by temperature $T$ and baryon
densities $\rho_i$.  A system is stable against separation into
two phases if its free energy is lower than the free energy in all
two-phase configurations.  This statement is formulated as
\begin{equation}
{\cal {F}}\left( T,\rho _i\right) <(1-\lambda) {\cal {F}}\left(
T,\rho _i'\right)+\lambda{\cal{F}}\left( T,\rho _i''\right)
\end{equation}
with
\begin{equation}
\rho _i=(1-\lambda) \rho _i'+\lambda\rho_i''.
\end{equation}
 The two phases
are denoted by a prime and a double prime.  In asymmetric nuclear
matter, we may choose $\rho_{i}$ as nuclear density $\rho$ and
asymmetry parameter $\alpha$.  The parameter $\lambda=V''/V$
specifies the volume fraction of the phase with double prime.
Eq.(18) ensures the overall baryon densities are conserved.  In
the SHM here, we may use variables $\rho_B,f_s$ instead of
$\rho_i$.  The above stability condition implies the following set
of inequalities[17]
\begin{equation}
\rho \left( \frac{\partial p}{\partial \rho_B}\right)_{T, f_s}
= \rho ^2\left( \frac{\partial ^2{\cal{F}}}{\partial
\rho_B^2}\right)_{T, f_s} >0
\end{equation}
\begin{equation}
\left( \frac{\partial \mu _N}{\partial f_s }\right) _{T,p}<0 \quad
or \quad \left( \frac{\partial \mu _\Xi}{\partial f_s }\right)
_{T,p}>0
\end{equation}
where $p$, $\mu_N$ and $\mu_\Xi$ are, respectively, the pressure,
chemical potentials for nucleon and $\Xi$.  The first inequality
Eq.(19) means that the isothermal compressibility is positive,
i.e. the system is mechanically stable.  The second condition
Eq.(20) reflects the special character of the binary system.  It
expresses "diffusive stability" which guarantees that energy is
required to change the concentration in a stable system, while
holding the remaining variables ($p$ and $T$) fixed.

Now we begin to discuss the region $f_s \leq 1.0$\cite{Schaffner2}
and show the possibility for an L-G phase transition in the SHM.
To be more specific, Fig.1 shows the pressure-baryon density
isotherms at temperature $T = 8 MeV$ with different strangeness
fractions ($f_s = 0.1, 0.3, 0.5, 0.8$ and $1.0$).  One can see
that there is always a mechanical unstable section where condition
(19) is violated for any strangeness fraction $f_s$ considered.
This means that the system always encounters an unstable region
and has to separate into two phases to maintain the mechanical
stability. This phenomenon remains up to $T \simeq 13 MeV$.  At
temperature $T = 13 MeV$, as is illustrated in Fig.2, there is a
mechanical inflection point which satisfies the condition
\begin{equation}
\frac{\partial p}{\partial \rho}|_{f_s=f_s^M} = \frac{\partial ^{2} p}{\partial \rho ^{2}}|_{f_s=f_s^M} = 0
\end{equation}
with $f_{s}^{M} = 0.5$ and the corresponding pressure $p = 0.24
MeV fm^{-3}$, $\rho_{B} = 0.063 fm^{-3}$.  For $fs \geq 0.8$, it
starts to decrease in high density region and mechanical unstable
section will occur. However, one cannot find the solution which
satisfies the Gibbs conditions in this region (See below in
Fig.7). To find the temperature dependence of $f_{s}^{M}$, we also
show the situation for $T = 14 MeV$ in Fig.3.  The mechanical
inflection point is around $p = 0.18 MeVfm^{-3}$ and $\rho_{B} =
0.04fm^{-3}$ on the isotherm with $f_{s} = 0.1$ for $T = 14 MeV$.
We find the $f_s^{M}$ decreases when the temperature increases. In
Fig.4, we present the pressure-density isotherms of fixed
strangeness fraction $f_{s} = 0.5$ at different temperatures.  One
can see again an inflection point on the curve with $T = 13 MeV$.

Next, we will discuss the chemical instability by showing the
chemical potential isobars for nucleons and $\Xi$ against
strangeness fractions $f_s$ at temperature $T=13MeV$ for pressure
$p=0.10,0.18,0.28$ and $0.35 MeVfm^{-3}$, respectively, in Fig.5.
 There is an inflection point on the curve with
pressure $p^C=0.28 MeVfm^{-3}$.  This pressure is called critical
pressure\cite{Muller, Qian3}, above which the system is stable but
below which condition (20) is violated and the system becomes
chemical unstable.  The critical pressure $p^C$ is determined by
the inflection point condition:
\begin{equation}
\left( \frac{\partial \mu _N}{\partial f_s }\right) _{T,p}=\left( \frac{%
\partial ^2\mu _N}{\partial f_s ^2}\right) _{T,p}=0.
\end{equation}
The result $(p^C, f_s^C)=(0.28MeVfm^{-3},0.65)$ defines a critical
point for a given temperature $T=13MeV$.

When the system separates into two phases, the two coexistent
phases are governed by the Gibbs conditions, which say that for
two phases (or more) the system should be in chemical, thermal and
mechanical equilibrium.
\begin{equation}
\mu _q^L\left( T,\rho^L, f_s^L\right) =\mu _q^G\left( T,\rho^G,
f_s^G\right), (q=N,\Xi),
\end{equation}
\begin{equation}
p^L\left( T,\rho^L,f_s^L\right) =p^G\left( T,\rho^G,f_s^G\right),
\end{equation}
where the superscripts $L$, $G$ denote the liquid and gas phases,
respectively.  Here we only consider the situation of two-phase
case.  As to more than two phases, there is some discussion in
ref.\cite{Muller}. The solution $\lbrace
\rho^L,f_s^L;\rho^G,f_s^G\rbrace$ specifying the two separate
phases in equilibrium can be easily found through a geometric
approach.  As an example, we present in Fig.6 the
chemical-strangeness fraction isotherms at temperature $T = 10
MeV$ and pressure $p = 0.09 MeVfm^{-3}$.  The desired solutions
form the edges of a rectangle shown by dotted lines. The phase
with lower(higher) $f_s$ corresponding to a higher(lower) density
is liquid(gas) phase.  It is obvious that the strangeness
fractions are different in the two phases.
 The collection of all such pairs ($p,T,f_s^L$) and ($p,T,f_s^G$)
forms a binodal surface of two dimensions, which defines the phase
separation boundary.  The shape of the whole binodal surface is
similar to that in Ref.\cite{Muller,Qian3}.  To explain clearly we
show in Fig.7 a section of the binodal surface at $T = 13 MeV$.
 In Fig.7 the binodal curve is divided into two
branches by zero strangeness point and critical pressure(CP)
point.  The branch for larger $f_{s}$ represents the gas phase
with lower density and the other for the liquid phase with higher
density.  In an isothermal compression, a system with zero
strangeness evolves until it encounters the binodal and then
remains there until the transition is completed.  In an isothermal
compression with $f_{s}\geq 0.8$, the L-G phase transition cannot
take place because the isothermal line does not cross the binodal
surface. At CP point, the liquid branch and gas branch joint
together smoothly and then the two phases can no longer be
distinguished by their densities.  The CP point determines the
maximum pressure in two-phase region.  If the system is
represented by the point in the section lower than the gas phase
branch, the system is in the gas phase, whereas if the system is
at the point higher than the liquid phase branch, it is in the
liquid phase.  Between the two branches, the system is in the
mixed phase with a special proportion between liquid and gas
phases.  The feature of the binodal curve at temperature $T = 13
MeV$ is very similar to those in asymmetric nuclear
matter\cite{Muller, Qian3}.  But for the temperature lower than $T
= 13 MeV$, the critical points could not be attained.
 Instead, we can always find a limit pressure in the binodal
surface.  As an example, we show in Fig.8 the section of the
binodal surface at the temperature $T = 10 MeV$, where a limit
pressure $p_{lim} = 0.095 MeVfm^{-3}$ is denoted by a horizontal
dotted line.  When pressure is higher than the limit pressure
$p_{lim}$, the L-G phase transition cannot take place. The cut-off
behavior for the binodal surface has also been found in asymmetric
nuclear matter described by FST model with density-dependent
$NN\rho$ coupling $g_{\rho}$\cite{Qian1, Qian3}. If $g_{\rho}$ is
constant, the cut-off behavior will not happen so that a critical
point could always be found and no limit pressure is obtained.  In
this case, the L-G phase transition becomes more complicated.  The
limit pressure exists when $T < 13 MeV$ even all couplings are
constant.

The Maxwell construction of the phase transition can narrate a
phase transition clearly and give a concrete proof for the
existence of the L-G phase transition.  For a given temperature,
we will expatiate the behavior of SHM under isothermal
compression.  At $T=10MeV$, for instance, Fig.8 illustrates the
situation of the isothermal compression.  Assume that the system
is initially in the gas phase with strangeness fraction $f_{s} =
0.15$.  During the compression, the phase boundary is encountered
at point A.  At this point, the liquid phase is about to emerge at
point B, where the strangeness fraction $f_{s}^{B}=0.05$.  During
the whole compression, the total strangeness fraction of the
system conserves but the strangeness fractions for the liquid and
gas phases are changeable.  As the compression proceeds, the gas
phase evolves along the gas phase branch of the binodal surface
from point A to D, while the liquid phase evolves from B to C
along the liquid branch.  When the pressure of the system
continues to increase, the system leaves the two-phase region at
point C, which has the same strangeness fraction with point A.
 Correspondingly, at point D on the gas branch with strangeness
fraction $f_s=0.4$, the gas phase disappears.  Since the two
points A and C are at different branches of the binodal surface,
the SHM has undergone an L-G phase transition.

To configure out the evolution of the system between point A and
C, we must solve the following conservation equations
\begin{equation}
\rho_{B}=(1-\lambda)\rho_{B}^{G} + \lambda\rho_{B}^{L},
\end{equation}
\begin{equation}
\rho_{B}f_{s}=(1-\lambda)\rho_{B}^{G}f_{s}^{G} +
\lambda\rho_{B}^{L}f_{s}^{L},
\end{equation}
for the baryon density, strangeness fractions on the binodal
surface and the proportion $\lambda$ for given total baryon
density $\rho_B$ and strangeness fraction $f_s$.  The proportion
$\lambda$ is $0$ at point A and $1$ at point C, and it runs
through the interval $[0, 1]$.  The result is the generalized
Maxwell construction in the binary system.  The isotherms
corresponding to Fig.8 are drawn in Fig.9.  The system dose not
evolve along the unphysical curve(the dotted line).  It follows
the nearly strait solid line between point A and C, which is the
result of the Maxwell construction.  Each point on the line
corresponds to the stable configuration at each intermediate
density during the phase transition.  Moreover, if the system is
initially in the gas phase with $f_s = 0.5$, the process of
isothermal compression will make that the system begins at the gas
phase, then enters a two phase region and becomes unstable at the
limit pressure $p_{lim}$.

Finally, we would like to discuss the order of the L-G phase
transition in the SHM.  By Ehrenfest's definition, the first order
phase transition is characterized by the discontinuities of the
first order derivatives of the chemical potential, such as the
discontinuities of entropy and volume, while the second order
phase transition unfolds the discontinuous behavior for the second
order derivatives of the chemical potential, such as the heat
capacity.
 Using the entropy density of the SHM in Eq.(16), we have the
entropy per baryon as
\begin{equation}
s(T,p,f_{s})=\frac{S(T,p,f_{s})}{\rho_{B}}.
\end{equation}
The total entropy per baryon of the system is calculated by the
equation:
\begin{equation}
s=(1-\lambda)s^{G} + \lambda s^{L}.
\end{equation}
The entropy per baryon as a function of temperature for a matter
of $f_{s} = 0.50$ during the phase transition at a fixed pressure
$p = 0.17 MeVfm^{-3}$ is delineated in Fig.10.  It is obvious that
the entropy per baryon varies continuously during the phase
transition.  To make our result more transparent, we calculate the
heat capacity $C_p$ of the SHM.
\begin{eqnarray}
C_{p} &=& T(\frac{\partial S}{\partial T})_{p,f_{s}}\nonumber \\
&&= T \sum_{i}\frac{g_{i}}{6\pi ^{2}}\int dk\frac{k^{4}}{
E_{i}^{\ast }(k)} \left[\frac{d^{2}n_{i}}{dT^{2}} +
\frac{d^{2}\overline{n}_{i}}{dT^{2}} \right],
\end{eqnarray}
where
\begin{eqnarray}
\frac{d^{2}n_{i}}{dT^{2}}= \frac{ e^{  \beta \left( E_{i}^{\ast
}(k)-\nu _{i}\right) }\left[( E_{i}^{\ast }(k)-\nu _{i})^{2}(e^{
\beta \left( E_{i}^{\ast }(k)-\nu _{i}\right) } - 1)-2T(
E_{i}^{\ast }(k)-\nu _{i}) (e^{\beta \left( E_{i}^{\ast }(k)-\nu
_{i}\right)}+1)\right]
 }{T^{4}(e^{\beta \left( E_{i}^{\ast
}(k)-\nu _{i}\right)}+1)^{3}},
\end{eqnarray}

\begin{eqnarray}
\frac{d^{2}\overline{n}_{i}}{dT^{2}}= \frac{ e^{  \beta \left(
E_{i}^{\ast }(k)+\nu _{i}\right) }\left[( E_{i}^{\ast }(k)+\nu
_{i})^{2}(e^{ \beta \left( E_{i}^{\ast }(k)+\nu _{i}\right) } -
1)-2T( E_{i}^{\ast }(k)+\nu _{i}) (e^{\beta \left( E_{i}^{\ast
}(k)+\nu _{i}\right)}+1)\right]
 }{T^{4}(e^{\beta \left( E_{i}^{\ast
}(k)+\nu _{i}\right)}+1)^{3}}.
\end{eqnarray}
The heat capacity per baryon is expressed as $c_{p} =
\frac{C_{p}}{\rho_{B}}$.  In Fig.11. the specific heat capacity as
a function of temperature for a fixed pressure $p = 0.17
MeVfm^{-3}$ for $f_{s} = 0.50$ is presented.
 A finite discontinuity of the heat capacity is
observed, which demonstrates clearly that the L-G phase transition
in SHM is of second order.

Till now, we can draw out a definite conclusion that an L-G phase
transition do exist in the SHM and it is of second order.

\section{Summary}
In this paper we have employed an extended FST model with nucleons
and $\Lambda$, $\Xi$ hyperons to describe the thermodynamical
properties of SHM.  By using the reactions between hyperons and
nucleons and making some assumptions, we have simplified the SHM
into a two-component system, which is mathematically feasible to
be discussed.  When having set the strangeness fraction to be
zero, the model can also obtain results, which match that of
symmetric nuclear matter. Applying the model, we have investigated
the L-G phase transition of the SHM. The mechanical unstable
region always exists for all strangeness fractions at the
temperatures lower than $T = 13 MeV$. Meanwhile, chemical
instability is also found in chemical potential-strangeness
fraction isobars at these temperatures.  By using the Gibbs
conditions of phase equilibrium, we have figured out the binodal
surface as the phase separation boundary and have found a limit
pressure at each temperature, above which there is no L-G phase
transition.  The limit pressure gives rise to a cut-off on the
section of the binodal surface. But for temperature as high as
$T=13MeV$, a critical pressure has been obtained in the binodal
surface section.  At the critical pressure point the liquid and
gas branches joint together, resulting in a closed binodal curve.
The Maxwell constructions for the phase transition have been
depicted in detail.  The entropy and heat capacity have also been
examined and the L-G phase transition in the SHM has been
determined to be of second order.

\section{Acknowledgements}
This work is supported in part by National Natural Science
Foundation of China under Nos. 10235030, 10247001, 10375013,
10347107 10047005, 10075071 by the National Basic Research
Programme 2003CB716300, by the Foundation of Education Ministry of
China under contract 2003246005 and CAS Knowledge Innovation
Project No. KJCX2-N11.

\begin{figure}[tbp]
\includegraphics[scale=0.5]{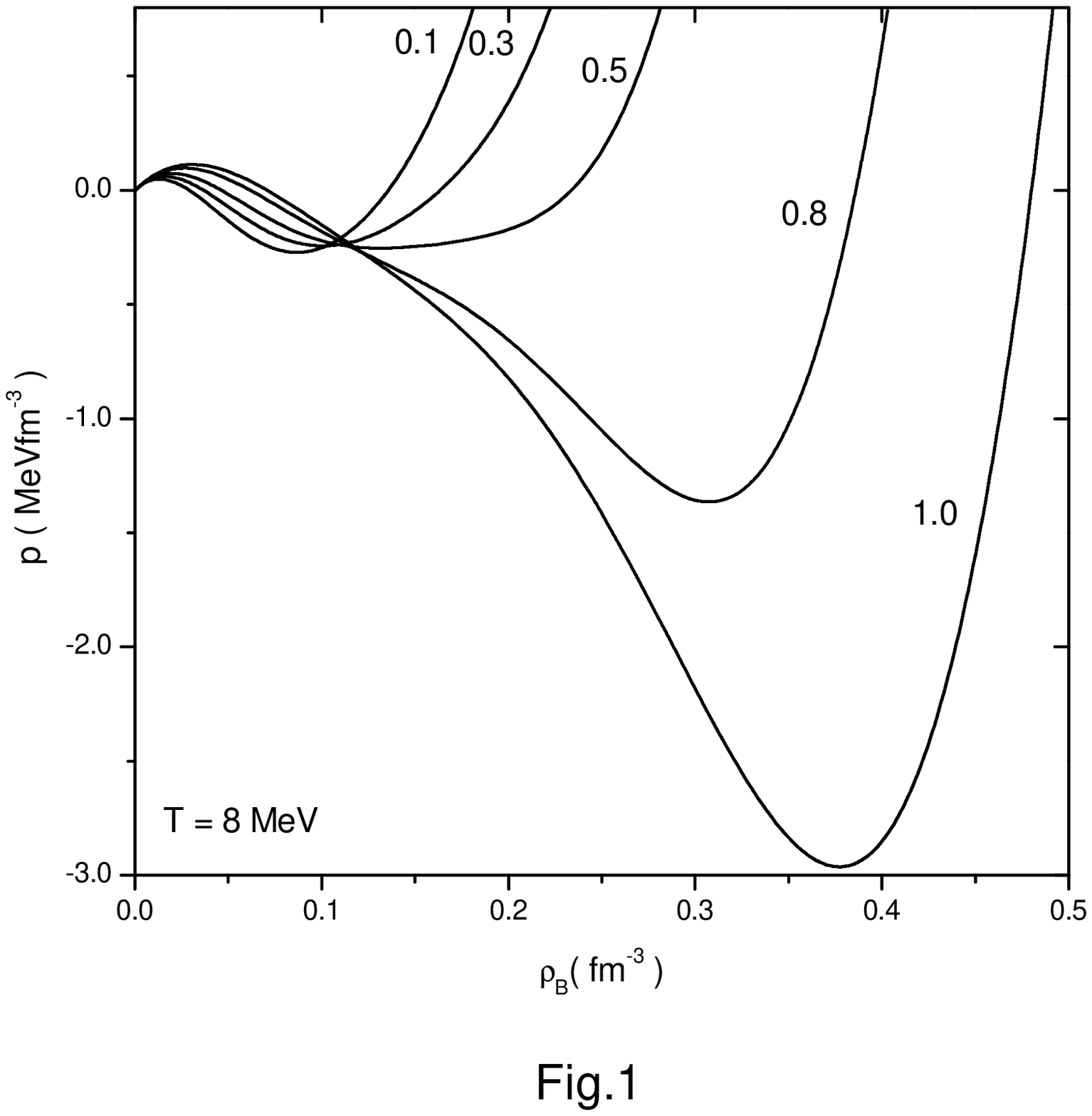}
\caption{Pressure as a function of baryon density at temperature
$T = 8 MeV$ for various strangeness fractions $f_{s}$.}
\label{fig1}
\end{figure}

\begin{figure}[tbp]
\includegraphics[scale=0.5]{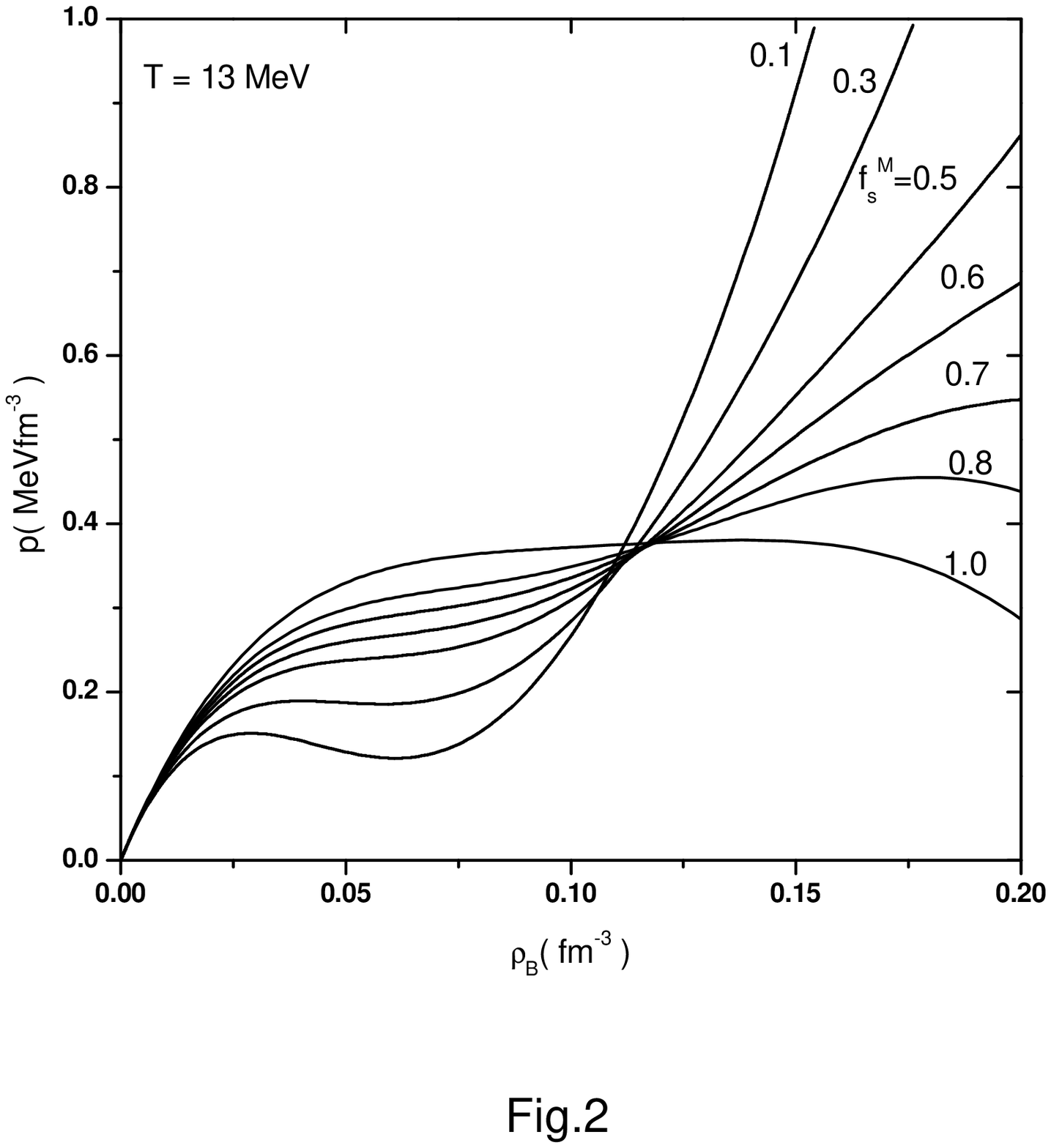}
\caption{Pressure as a function of baryon density at temperature
$T = 13 MeV$ for various strangeness fractions $f_{s}$. The
mechanical inflection point is at isotherm with $f_{s}^{M} =
0.5$.} \label{fig2}
\end{figure}

\begin{figure}[tbp]
\includegraphics[scale=0.5]{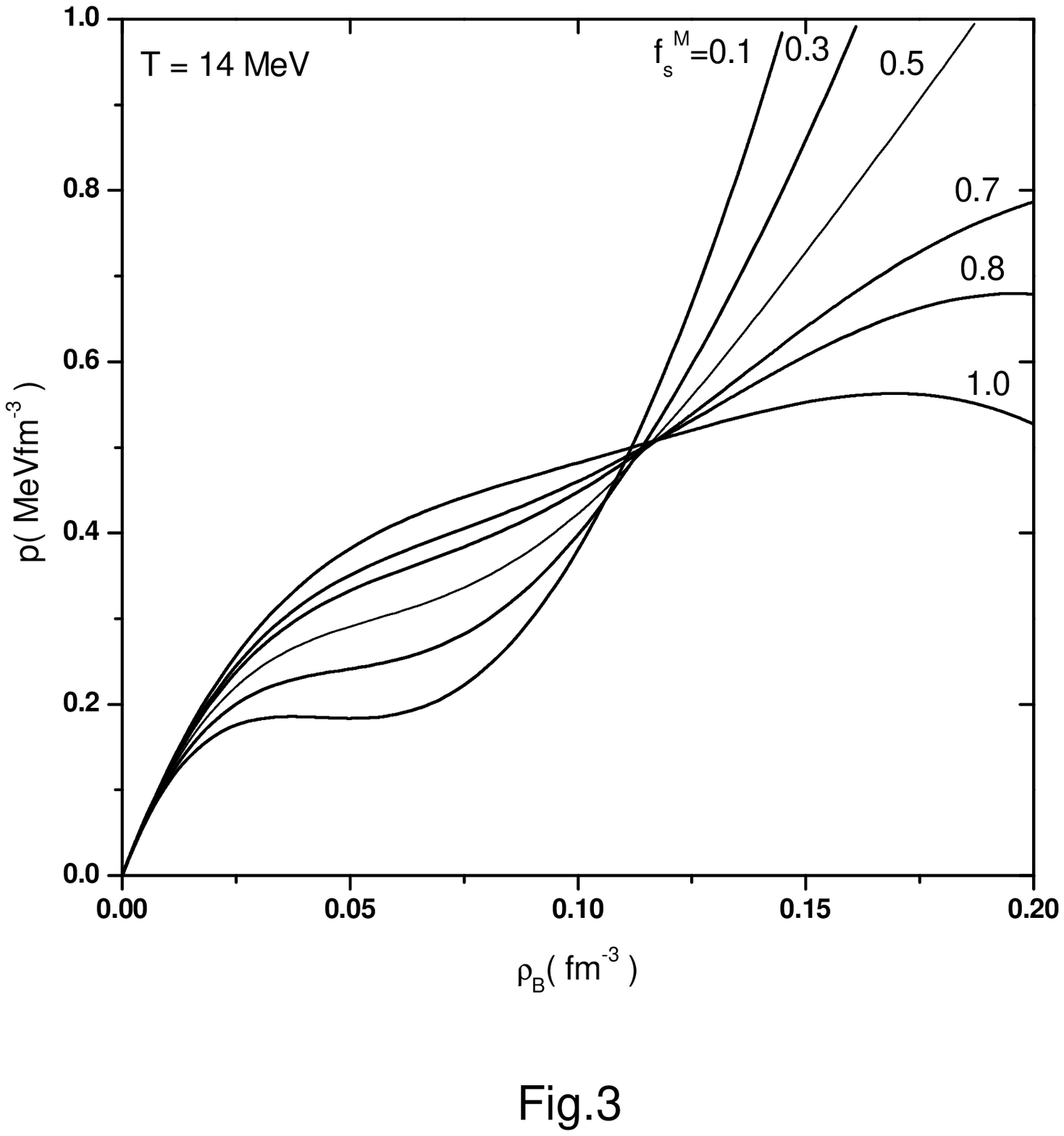}
\caption{Pressure as a function of baryon density at temperature
$T = 14 MeV$ for various strangeness fractions $f_{s}$. A
mechanical inflection point is found at isotherm with $f_{s}^{M} =
0.1$.} \label{fig3}
\end{figure}

\begin{figure}[tbp]
\includegraphics[scale=0.5]{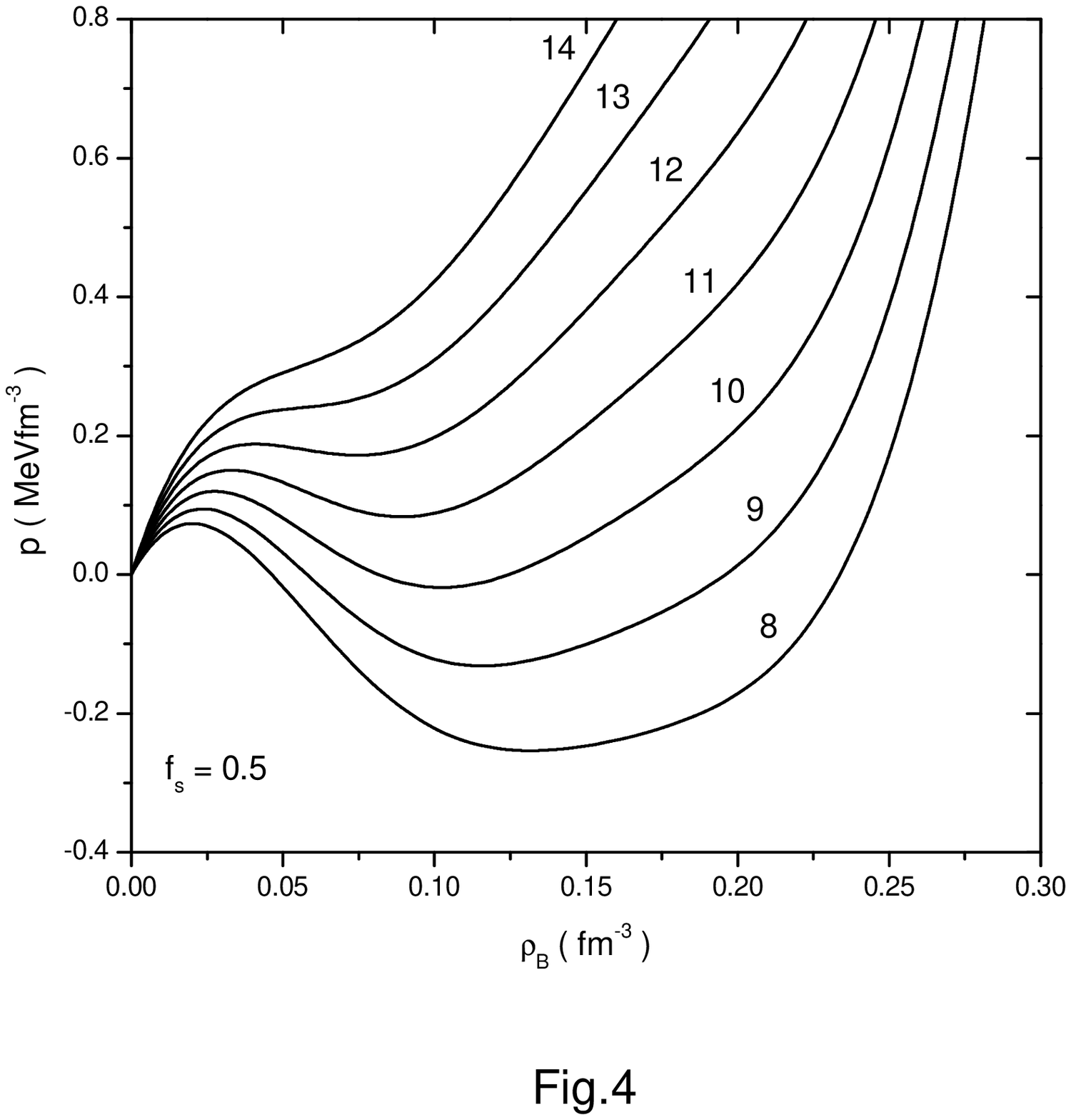}
\caption{Pressure as a function of baryon density for fixed
strangeness fraction $f_{s} = 0.5$ at various temperature. A
mechanical inflection point is at isotherm $T=13MeV$, below which,
isotherms have the mechanical unstable regions.} \label{fig4}
\end{figure}

\begin{figure}[tbp]
\includegraphics[scale=0.5]{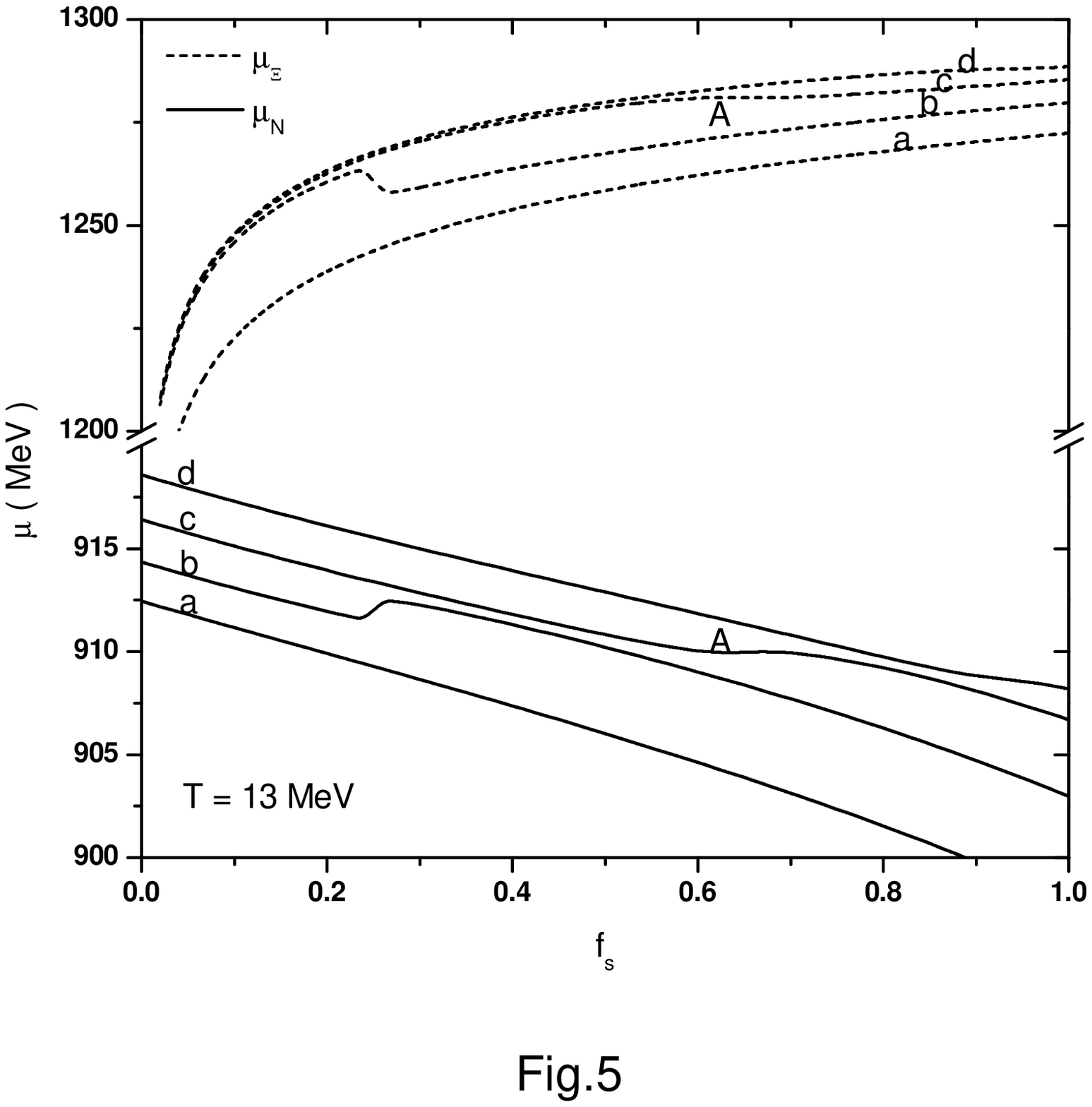}
\caption{Chemical potential isobars as a function of the
strangeness fraction $f_{s}$ at fixed temperature $T = 13 MeV$.
The curves labeled a through d have pressures $p = 0.10, 0.18,
0.28$ and $0.35 MeV fm^{-3}$ respectively. Point A denotes the
inflection point at the critical pressure $p^{C}=0.28MeVfm^{-3}$.}
\label{fig5}
\end{figure}

\begin{figure}[tbp]
\includegraphics[scale=0.5]{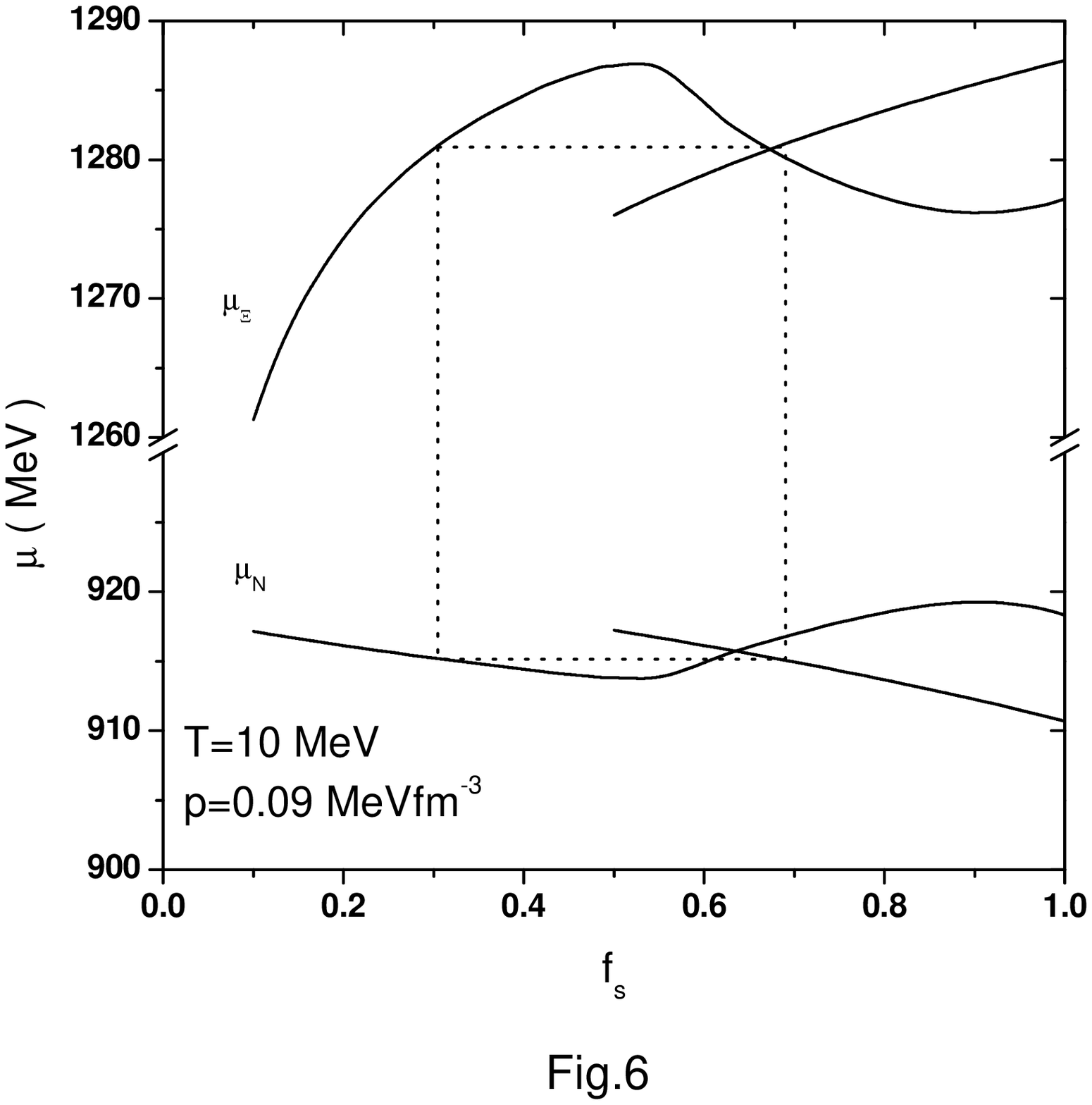}
\caption{Geometrical construction used to obtain the strangeness
fraction and chemical potentials in the two coexisting phases at
fixed temperature $T = 10 MeV$ and pressure $p = 0.09MeVfm^{-3}$.
The two solid crossing curves for $\mu_{N}$ (or $\mu_{\Xi}$) are
the different parts of the same continuous curve.} \label{fig6}
\end{figure}

\begin{figure}[tbp]
\includegraphics[scale=0.5]{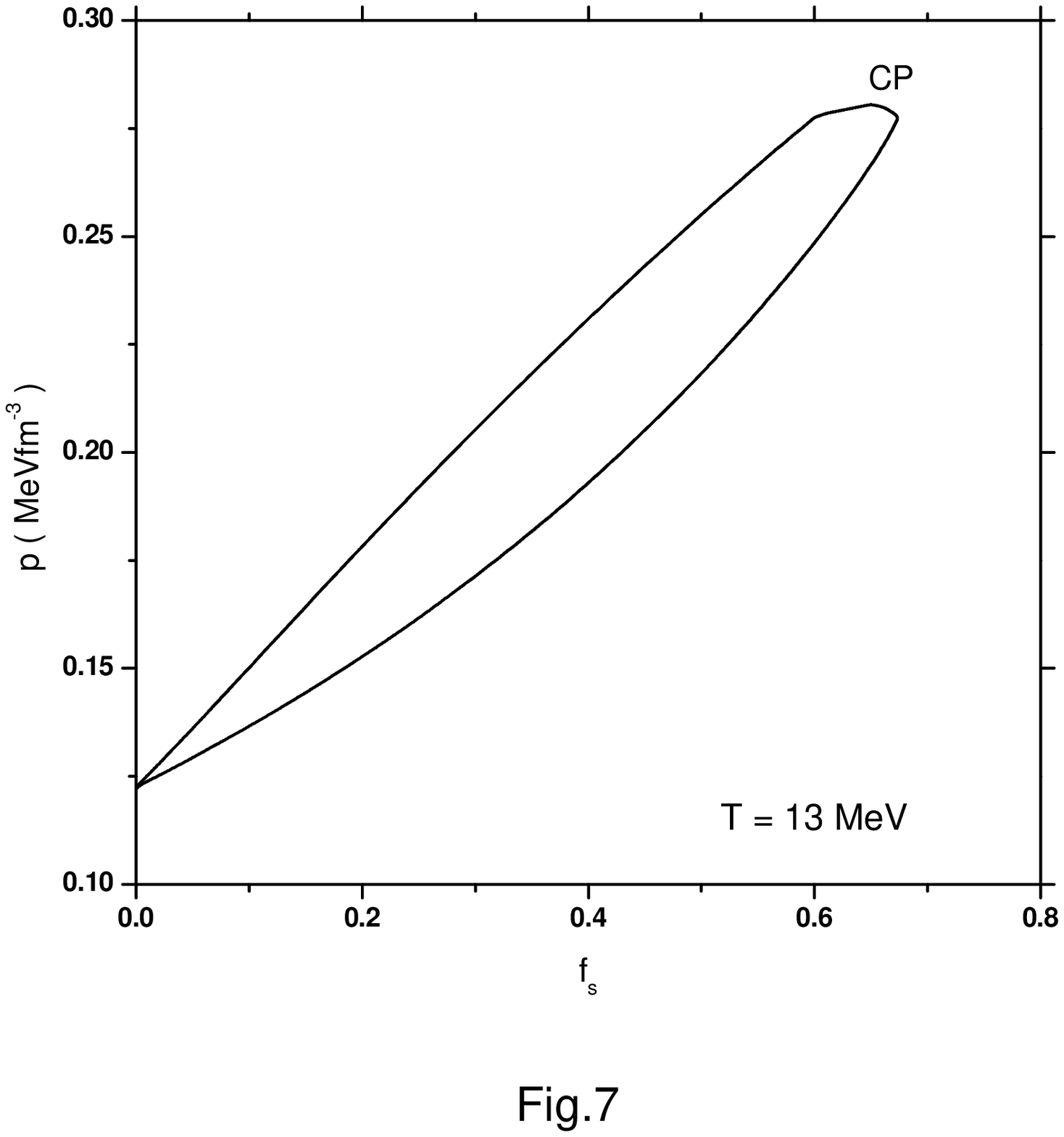}
\caption{Binodal surface at $T = 13 MeV$. A critical pressure(CP)
and the zero strangeness are indicated. The upper branch is for
liquid phase and the lower for gas phase. Two branches joint at CP
point.} \label{fig7}
\end{figure}

%\begin{figure}[tbp]
%\includegraphics[scale=0.5]{Fig8.eps}
%\caption{Binodal surface at $T = 10 MeV$. The binodal surface is
%cut off at a limit pressure $p = 0.095 MeVfm^{-3}$.} \label{fig8}
%\end{figure}

\begin{figure}[tbp]
\includegraphics[scale=0.5]{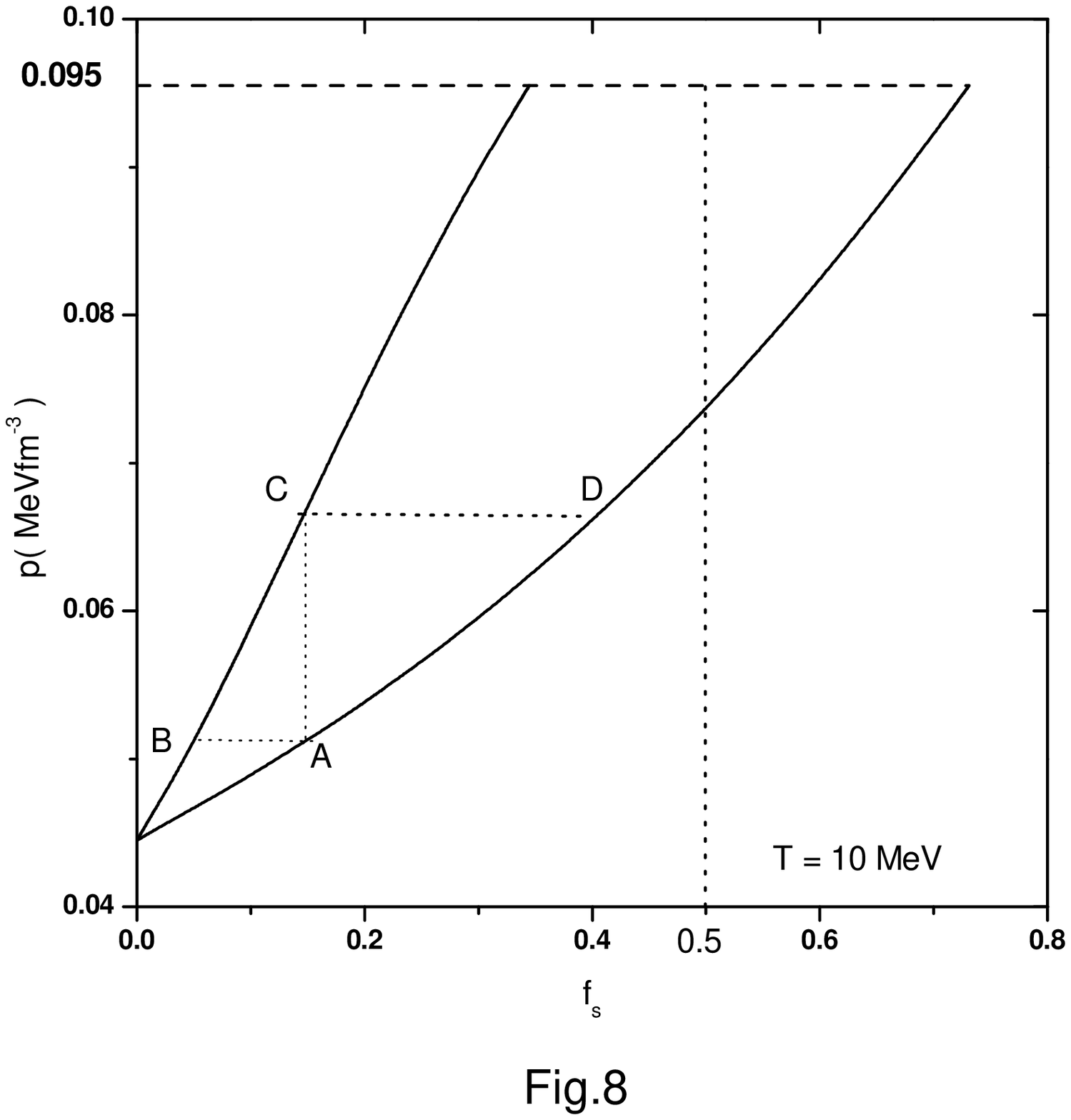}
\caption{Binodal surface at $T=10MeV$.  The binodal surface is cut
off at a limit pressure $p = 0.095 MeVfm^{-3}$.  The point A
through D denote phases participating in a normal phase
transition. Another phase transition for $f_{s}=0.5$ is also
illustrated by the dotted line.} \label{fig8}
\end{figure}

\begin{figure}[tbp]
\includegraphics[scale=0.5]{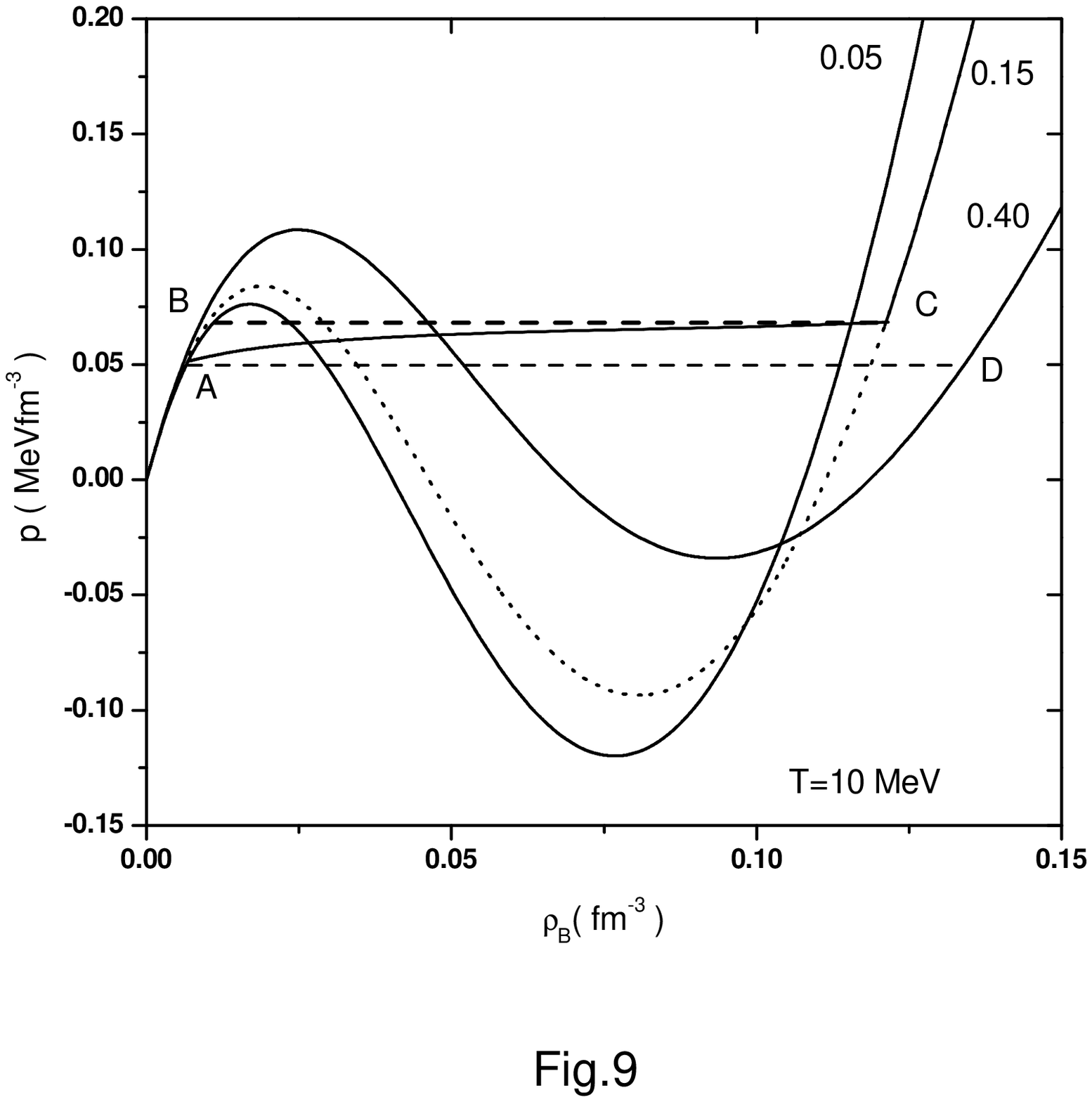}
\caption{Isotherms for a phase transition at $T=10MeV$ and the
initial condition $f_{s} = 0.15$. The solid line AC is built up by
Maxwell construction, and the system evolves along AC.}
\label{fig9}
\end{figure}

\begin{figure}[tbp]
\includegraphics[scale=0.5]{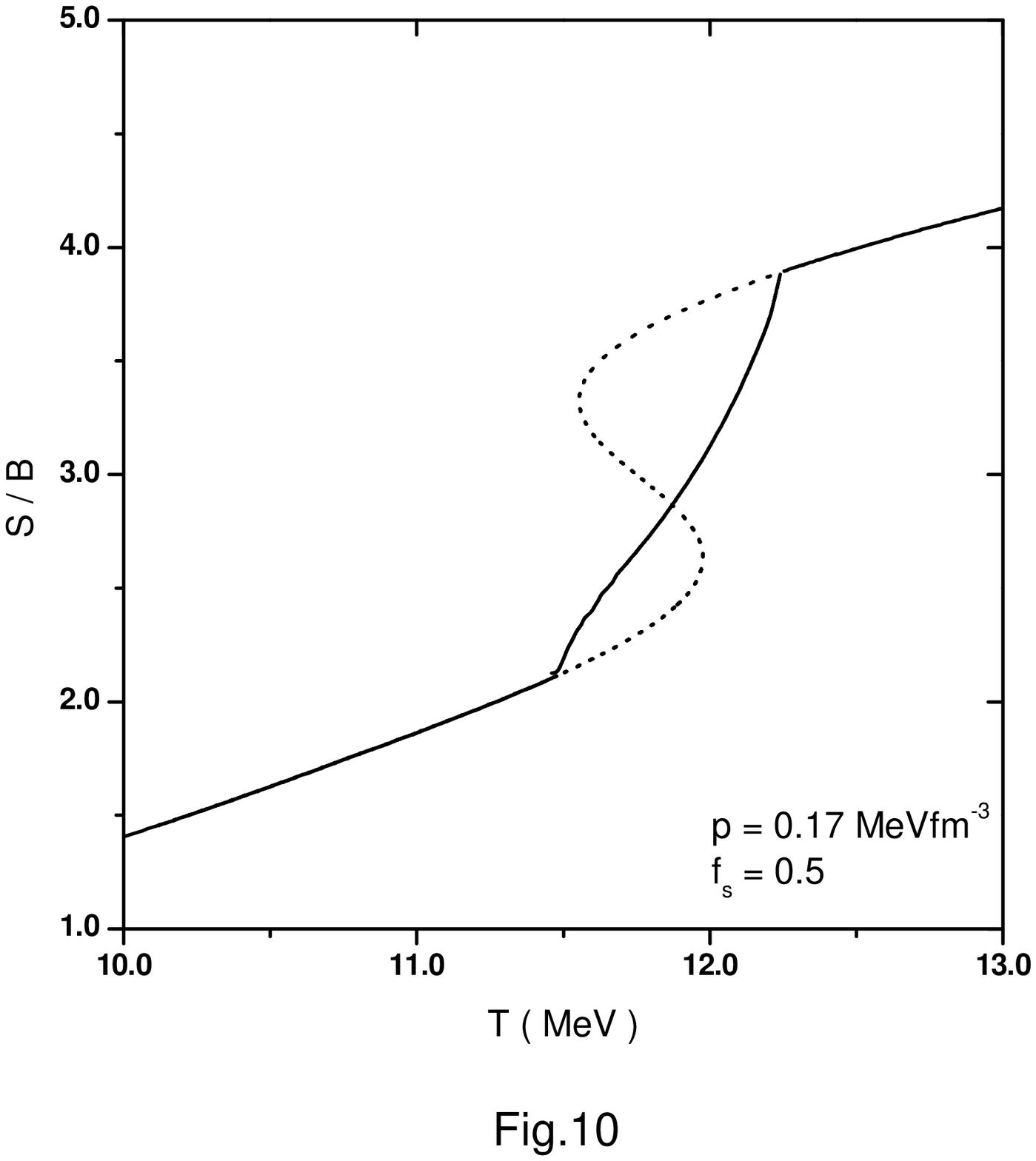}
\caption{Entropy as a function of temperature at constant pressure
$p=0.17MeVfm^{-3}$ for strangeness fraction $f_s=0.50$. The
entropy evolves continuously through the phase transition.}
\label{fig10}
\end{figure}

\begin{figure}[tbp]
\includegraphics[scale=0.5]{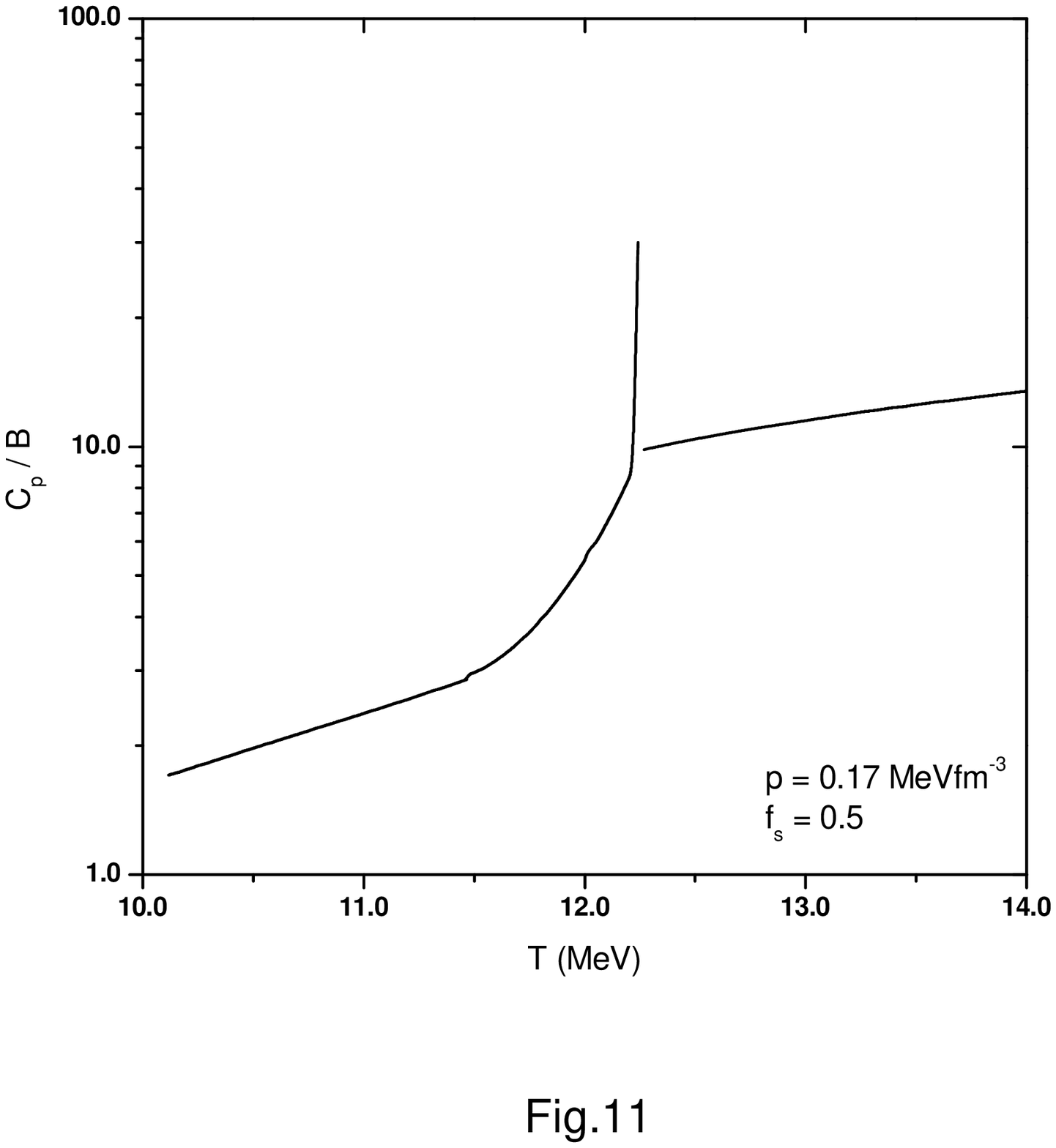}
\caption{Heat capacity as a function of temperature at fixed
pressure $p=0.17MeVfm^{-3}$ for strangeness fraction $f_s=0.50$.
The discontinuity makes a strong proof for the second order phase
transition. Note the logarithmic scale used on the left.}
\label{fig11}
\end{figure}


\begin{references}
\bibitem{Kupper}W.A. K\"upper, G. Wegmann and E.R. Hilf, Ann. Phys. {\bf 88},454(1972).
\bibitem{Lamb}D.Q. Lamb et al., Phys. Rev. Lett.{\bf 41},1623(1978).
\bibitem{Bertsch} G. Bertsch and P. J. Siemens Phys. Lett. {\bf B126},9 (1983).
\bibitem{Curtin}M.W. Curtin, H. Toki and D.K. Scott, Phys. Lett. {\bf B123},289(1983).
\bibitem{Finn}J.E. Finn, et al., Phys. Rev. Lett.{\bf 49},1321(1982).
\bibitem{Jaqaman}H.R. Jaqaman, A.Z. Mekjian, and L. Zamick, Phys. Rev. {\bf C27},2782(1983);{\bf 29},2067(1984).
\bibitem{Su}R.K. Su, S.D. Yang, and T.T.S. Kuo, Phys. Rev. {\bf C35},1539(1987).
\bibitem{Satpathy}L. Satpathy, M. Mishra, and R. Nayak, Phys. Rev. {\bf C39},162(1989).
\bibitem{Kapusta}J. Kapusta, Phys. Rev. {\bf C29},1735(1984).
\bibitem{Goodman}A.L. Goodman, J.I. Kapusta, and A.Z. Mekjian, Phys. Rev. {\bf C30},851(1984).
\bibitem{Boal}D.H. Boal and A.L. Goodman, Phys. Rev. {\bf C33},1690(1986).
\bibitem{Siemens}P.J. Siemens, Nature(london){\bf 305}, 410(1983).
\bibitem{EOS}J.B. Elliott et al., Phys. Rev. {\bf C49},3185(1994); M.L. Gilkes et al., Phys. Rev. Lett. {\bf 73},1590(1994).
\bibitem{ALADIN}J. Pochodzalla et al., Phys. Rev. Lett. {\bf 75},1040(1995).
\bibitem{Agostino} M.D'Agostino, M.Bruno, F.Gulminelli, R.Bougault, F.Cannata, Ph.Chomaz, F.Gramegna, N.LeNeindre, A.Moroni, G.Vannini, Proceedings of the 8th International Conference on Nucleus-Nucleus Collisions, Moscow 2003;  Nucl. Phys. {\bf A734},512(2004).
\bibitem{nucl-th/0305050} Takuya Furuta, Akira Ono, nucl-th/0305050
\bibitem{Muller} H. M\"{u}ller and B. D. Serot, Phys. Rev. {\bf C52}, 2072 (1995).
\bibitem{Qian1} W. L. Qian, R. K. Su and P. Wang, Phys. Lett. {\bf B491},90 (2000).
\bibitem{Qian2} W. L. Qian, R. K. Su and H. Q. Song, Phys. Lett. {\bf B520},217 (2001).
\bibitem{Qian3} W. L. Qian, R. K. Su and H. Q. Song, J. Phys. {\bf G28}, 379 (2002).
\bibitem{Qian4} W. L. Qian, R. K. Su, J. Phys. {\bf G29}, 1023 (2003).
\bibitem{Kolomietz} V. M. Kolomietz, A. I. Sanzhur, S. Shlomo and S. A. Firin, Phys. Rev. {\bf C64}, 024315 (2001).
\bibitem{Karnaukhov} V. A. Karnaukhov and et. al., Phys. Rev. {\bf C67},011601 (2003).
\bibitem{Pawlowski} P. Pawlowski Phys. Rev. {\bf C65},044615 (2002).
\bibitem{Danysz}M. Danysz and J. Pniewski, {\it Phil. Mag.} {\bf 44}, 348 (1953).
\bibitem{Ikeda} K. Ikeda, H. Bando and T. Motoba, Prog. Theor. Phys. Supplement {\bf 81}, 47 (1995).
\bibitem{Schaffner1} J. Schaffner, C. B. Dover, A. Gal, C. Greiner, D. J. Millener and H. Stocker, Ann. Phys. {\bf 235} (1994) 35
\bibitem{Schaffner2} J. Schaffner-Bielich and A. Gal, Phys. Rev. {\bf C62}, 034311 (2000).
\bibitem{Schulze} H. J. Schulze, M. Baldo, U. Lombardo, J. Cugnon and A. Lejeune, Phys. Rev. {\bf C57}, 704 (1998).
\bibitem{Lee} Y. Zhang and R. K. Su, Phys. Rev. {\bf C65}, 035202(2002), Phys. Rev. {\bf C67},
015202(2003)
%\bibitem{Lee} K. S. Lee and U. Heinz, Phys. Rev. {\bf D47}, 2068 (1993).
\bibitem{Zhang1} L. L. Zhang, H. Q. Song, P. Wang and R. K. Su  J. Phys. {\bf G26}, 2301 (2000).
\bibitem{Qian5} W. L. Qian, R. K. Su and H. Q. Song, Commun. Theor. Phys. {\bf 40} 466 (2003).
\bibitem{Furnstahl1997} R. J. Furnstahl, Hua-Bin Tang and Brian D. Serot, Nucl. Phys. {\bf A615}, 441(1997).
\bibitem{Furnstahl1995} R. J. Furnstahl, Hua-Bin Tang and Brian D. Serot, Phys. Rev. {\bf C52}, 1368(1995).
\bibitem{Mares}J. Mares, E. Triedman, A. Gal and B.K. Jennings, Nucl. Phys. {\bf A594},311(1995).
\bibitem{Dover}C.B. Dover, D.J. Millener and A. Gal, Phys. Rep.{\bf 184}, 1(1989).
\bibitem{Balberg}S. Balberg, A. Gal and J. Scharffner, Progr. Theor. Phys. Suppl.{\bf 117}, 325(1994).
\bibitem{Stoks}V.G.J. Stoks and T.-S.H. Lee, Phys. Rev. {\bf C60}, 024006-1(1999).
\bibitem{Lee2} S.J. Lee, A.Z. Mekjian, Phys. Rev. {\bf C56}, 2621 (1997).
\end{references}
\end{document}